\documentclass[11pt,a4paper]{article}
\pdfoutput=1
\usepackage{jcappub}
\usepackage[T1]{fontenc}
\title{Naturally Light Dirac Neutrino in Left-Right Symmetric Model}

\author[a]{Debasish Borah, \note{Corresponding author}}
\affiliation[a,1]{Department of Physics, Indian Institute of Technology Guwahati, Assam-781039, India}
\emailAdd{dborah@iitg.ernet.in}
\author[b]{Arnab Dasgupta}
\affiliation[b]{Institute of Physics, HBNI, Sachivalaya Marg, Bhubaneshwar-751005, India}
\emailAdd{arnab.d@iopb.res.in}

\abstract{
We study the possibility of generating tiny Dirac masses of neutrinos in Left-Right Symmetric Model (LRSM) without requiring the existence of any additional symmetries. The charged fermions acquire masses through a universal seesaw mechanism due to the presence of additional vector like fermions. The neutrinos acquire a one-loop Dirac mass from the same additional vector like charged leptons without requiring any additional discrete symmetries. The model can also be extended by an additional $Z_2$ symmetry in order to have a scotogenic version of this scenario predicting a stable dark matter candidate. We show that the latest Planck upper bound on the effective number of relativistic degrees of freedom $N_{\text{eff}}=3.15 \pm 0.23$ tightly constrains the right sector gauge boson masses to be heavier than 3.548 TeV. This bound on gauge boson mass also affects the allowed values of right scalar doublet dark matter mass from the requirement of satisfying the Planck bound on dark matter relic abundance. We also discuss the possible implications of such a scenario in charged lepton flavour violation and generating observable electric dipole moment of leptons.}
\begin{document}
\maketitle

\section{Introduction}
The fundamental origin of tiny neutrino masses and large leptonic mixing has been one of the most important motivations for studying beyond standard model (BSM) physics. Due to the absence of the right handed neutrinos in the standard model (SM), the neutrinos can not couple to the Higgs field at renormalisable level and hence remain massless thereby giving rise to zero leptonic mixing. However, vanishing neutrino mass or vanishing leptonic mixing has already been ruled out by several experimental observations \cite{PDG, kamland08, T2K, chooz, daya, reno, minos} which suggest sub-eV scale neutrino mass and large leptonic mixing. The present status of different neutrino parameters can be found in the latest global fit analysis \cite{schwetz16}. These experiments have very precisely measured two mass squared differences and three mixing angle upto some ambiguity in the octant of the atmospheric mixing angle $\theta_{23}$ \footnote{See the latest experimental result from NovA collaboration on this \cite{NovA17}}. This keeps the lightest neutrino mass as well as the neutrino mass hierarchy undetermined at the present experiments. On the other hand, experiments at cosmic frontiers like the Planck can put an upper bound on the mass of the lightest neutrino from their bounds on the sum of absolute neutrino masses $\sum_i \lvert m_i \rvert < 0.17$ eV \cite{Planck15}. Apart from the lightest neutrino mass, the leptonic CP violating phase also is not determined conclusively at the neutrino experiments although there was a recent hint from the T2K experiment suggesting $\delta \approx -\pi/2$ \cite{diracphase}. 

The neutrino experiments mentioned above also remain insensitive to the Dirac or Majorana nature of neutrinos. If neutrinos are Majorana fermions, two more CP phases appear in the mixing matrix. But their effects or the consequences of the Majorana nature of neutrinos can be observed only at experiments like the ones looking for neutrinoless double beta decay experiments $(0\nu \beta \beta)$. Since a Majorana fermion is its own antiparticle, its very existence implies lepton number violation (LNV) in the neutrino sector. This is in fact, a generic feature of most of the BSM frameworks proposed so far in order to explain tiny but non-zero neutrino masses. The typical seesaw mechanisms \cite{ti, tii0, tii, tiii} of neutrino mass generation involves the introduction of heavy fermion or scalar fields to the SM which violates lepton number and give rise to light Majorana neutrinos having sub-eV mass. Without taking the route of such seesaw mechanism, if one introduces three copies of right handed neutrinos and allow a Dirac Yukawa term of the type $Y_{\nu} \bar{L} H \nu_R$, it can not generate a sub-eV Dirac mass naturally. This is because the dimensionless Yukawa coupling $Y_{\nu}$ has to be fine tuned to $~10^{-12}$ in order to generate a Dirac mass of 0.1 eV. Also, the symmetry of the SM can not prevent a Majorana mass term of the singlet right handed neutrino $\nu_R$ resulting in Majorana or pseudo-Dirac nature of light sub-eV neutrinos. Thus, to have purely Dirac type neutrinos, one usually has to introduce additional symmetries that can prevent a Majorana mass term of the heavy fields introduced for seesaw and also can explain the origin of tiny Dirac mass. There have been several proposals already that can generate tiny Dirac neutrino masses \cite{balamoha, babuhe, diracmass, diracmass1, ma1, ma2, ma3, db1, db2, db3}. While most of these scenarios explain the origin of tiny Dirac mass at one or two loop level, there are some scenarios \cite{diracmass1} which consider an additional scalar doublet apart from the SM one which acquire a tiny vacuum expectation value (vev) naturally due to the presence of a softly broken global symmetry. The radiative Dirac neutrino mass models also incorporate additional symmetries like $U(1)_{B-L}, Z_N$ in order to generate a tiny neutrino mass of purely Dirac type.

Here we study a simpler way of generating tiny neutrino masses of Dirac type without imposing any additional symmetries. Instead of considering the addition of singlet right handed neutrinos to the SM, here we consider another model where the presence of the right handed neutrinos is a necessity and not a choice. More specifically, we consider a specific realisation of the left-right symmetric models (LRSM), a popular BSM framework proposed several decades ago \cite{lrsm}. In these models, the gauge symmetry of the SM is enhanced to $SU(3)_c \times SU(2)_L \times SU(2)_R \times U(1)_{B-L}$ such that the right handed fermions transforming as $SU(2)_L$ singlets in the SM become doublet under $SU(2)_R$. Therefore, the right handed neutrino is automatically included into the model as a part of the right handed lepton doublet. The minimal version of this model gives rise to Majorana type light neutrinos through one of the seesaw mechanisms. A specific version of LRSM discussed in a much earlier work \cite{babuhe} generates Dirac neutrino mass at two loop level. Recently, there have been a few proposals to realise a tiny Dirac type neutrino mass at one loop level within different variants of LRSM \cite{ma3, db1, db3} where all of them considered additional discrete symmetries apart from the gauge symmetry mentioned above \footnote{At this point we note that, there have also been some proposals within LRSM which generate Majorana neutrino masses at one-loop \cite{db1, pavel1} and two-loop level \cite{radiativeLR}.}. Here we consider a similar scenario but without any discrete symmetries. This possibility was first pointed out by \cite{ma1989} which we study in details in this work. We incorporate an additional $Z_2$ discrete symmetry only when we take dark matter into account. This follows the basic idea of scotogenic models first proposed by \cite{ma06}. The scotogenic version of an LRSM for Dirac neutrinos was recently proposed by the authors of \cite{ma3} but with $Z_2 \times Z_2$ as additional discrete symmetries and introducing heavier charged fermions in the $SU(2)_R$ doublet representations instead of the usual right handed charged fermions of the SM. Two other versions of scotogenic LRSM was proposed in \cite{db1}, \cite{db3} with $Z_4$ and $Z_4 \times Z_4$ as additional discrete symmetries respectively. Here we study a simpler and more minimal version of \cite{db1,db3} where we can generate tiny Dirac neutrino masses in scotogenic fashion just with $Z_2$ as an additional discrete symmetry. 

In both the models we discuss in this work with and without dark matter, the charged fermions acquire their masses from a universal seesaw mechanism \cite{VLQlr, univSeesawLR} due to the presence of additional heavy fermions. Due to their vector like nature, the bare mass terms of these heavy fermions can be written in the Lagrangian. The charged fermion mass hierarchies can arise due to hierarchies in these heavy fermion masses. These models can also have interesting implications for cosmology. Due to the Dirac nature of light neutrinos, the effective number of relativistic degrees of freedom $N_{\text{eff}}$ is doubled from the scenarios where the light neutrinos are of Majorana type. In order to satisfy the the Planck upper bound on $N_{\text{eff}}$ \cite{Planck15} we find that the right handed gauge interactions which affect the decoupling of the right handed neutrinos must be suppressed giving a lower bound on the respective gauge boson mass $M_{W_R} > 8.387$ TeV. We show that such a bound also affects the relic abundance of right scalar doublet dark matter. Apart from these interesting cosmological implications, the model can also have interesting signatures in the flavour sector. The heavy charged leptons can also take part in lepton flavour violating decays like $\mu \rightarrow e \gamma$ as well as generating electric dipole moments (EDM) of charged leptons that can be looked for at rare decay experiments like MEG \cite{MEG16} or experiments measuring the EDM of charged leptons \cite{EDM2, EDM3, EDM4}. Also, since the present model predicts purely Dirac neutrinos, any observations of neutrinoless double beta decay $(0\nu \beta \beta)$ at present and future experiments \cite{kamland, kamland2, GERDA, GERDA2, exo200, 0nbbexpt} will be able to falsify such a scenario. Such light Dirac neutrinos could also have interesting implications in cosmology.

This paper is organised as follows. In section \ref{sec1}, we briefly discuss the minimal model with Dirac neutrino mass and set the motivations for its extension by a pair of scalar doublets discussed in section \ref{sec1b}. We then discuss the scotogenic extension of it in section \ref{sec2} followed by the discussion on cosmological implications of these models in section \ref{sec2b}. In section \ref{sec3} we briefly discuss different possible phenomenology related to flavour and collider physics and finally conclude in section \ref{sec4}.

\section{The Minimal Model}
\label{sec1}
The minimal LRSM without the Higgs bidoublet have been discussed extensively in the literature \cite{VLQlr, univSeesawLR}. The particle content of the model is shown in table \ref{tab:data1} and \ref{tab:data2}. The model is also an extension of the SM gauge symmetry to $SU(3)_c \times SU(2)_L \times SU(2)_R \times U(1)_{B-L}$ with the right handed fermions being doublets under $SU(2)_R$ similar to the way left handed fermions transform as doublets under $SU(2)_L$. The absence of the Higgs bidoublet prevents any tree level Yukawa couplings between the left and right handed fermions. However, addition of vector like heavy singlet fermions can generate the masses of all the known fermions through a universal seesaw mechanism. In the fermion content shown in table \ref{tab:data1}, the doublets are the usual LRSM fermion doublets and the vector like fermions $U, D, E$ are required for the universal seesaw for charged fermion masses. The scalar content shown in table \ref{tab:data2} are chosen in a way to allow Yukawa couplings between the fermion doublets and heavy fermion singlets. It should be noted that we are not including the neutral heavy leptons $N_{L,R}$ that are usually introduced in this model in order to generate tiny Majorana neutrino masses through universal seesaw. Since here we intend to discuss minimal ways of generating Dirac neutrino masses, we do not include such neutral leptons which lead to lepton number violation and hence to Majorana neutrinos. 
\begin{table}
\begin{center}
\begin{tabular}{|c|c|}
\hline
Particles & $SU(3)_c \times SU(2)_L \times SU(2)_R \times U(1)_{B-L} $   \\
\hline
$q_L=\begin{pmatrix}u_{L}\\
d_{L}\end{pmatrix}$ & $(3, 2, 1, \frac{1}{3})$  \\
$q_R=\begin{pmatrix}u_{R}\\
d_{R}\end{pmatrix}$ & $(3, 1, 2, \frac{1}{3})$  \\
$\ell_L=\begin{pmatrix}\nu_{L}\\
e_{L}\end{pmatrix}$ & $(1, 2, 1, -1)$  \\
$\ell_R=\begin{pmatrix}\nu_{R}\\
e_{R}\end{pmatrix}$ & $(1, 2, 1, -1)$ \\
$U_{L,R}$ & $(3, 1, 1, \frac{4}{3})$  \\
$D_{L,R}$ & $(3, 1, 1, -\frac{2}{3})$  \\
$ E_{L,R}$ & $(1,1,1, -2)$  \\
\hline
\end{tabular}
\end{center}
\caption{Fermion Content of the Model}
\label{tab:data1}
\end{table}
\begin{table}
\begin{center}
\begin{tabular}{|c|c|}
\hline
Particles & $SU(3)_c \times SU(2)_L \times SU(2)_R \times U(1)_{B-L} $   \\
\hline
$H_{L}=\begin{pmatrix}H^+_{L}\\
H^0_{L}\end{pmatrix}$ & $(1,2,1,-1)$  \\
$H_{R}=\begin{pmatrix}H^+_{R}\\
H^0_{R}\end{pmatrix}$ & $(1,1,2,-1)$  \\
\hline
\end{tabular}
\end{center}
\caption{Scalar content of the Model}
\label{tab:data2}
\end{table}

The Lagrangian for fermions can be written as
\begin{align}
\mathcal{L} & \supset Y_{U} (\overline{q_L} H_{L} U_L+\overline{q_R} H_{R} U_R) + Y_{D} (\overline{q_L} H^{\dagger}_{L} D_L+\overline{q_R} H^{\dagger}_{R} D_R) +M_U \overline{U_L} U_R+ M_D \overline{D_L} D_R\nonumber \\
& +Y_{E} (\overline{\ell_L} H^{\dagger}_{L} E_L+\overline{\ell_R} H^{\dagger}_{R} E_R) +M_E \overline{E_L} E_R+\text{h.c.}
\label{fermionL0}
\end{align}
The usual left and right handed fermions transforming as doublets under $SU(2)_{L,R}$ do not couple to each other directly at tree level due to the absence of the scalar bidoublet. But since they couple to the heavy vector like singlet fermions, one can generate their effective masses by integrating out the heavy fermions. The resulting charged fermion masses are given by 
$$m_u =  Y_{U} \frac{1}{M_U} Y^T_{U} v_{L} v_{R}, \;\; m_d = Y_{D} \frac{1}{M_D} Y^T_{D} v_{L} v_{R}, \;\; m_e =  Y_{E} \frac{1}{M_E} Y^T_{E} v_{L} v_{R} $$
where $v_{L,R} = \langle H^0_{L,R} \rangle$ are the vev's of the neutral components of the scalar fields. The gauge boson masses can be derived as
$$ M_{W_L} = \frac{g}{2}v_{L}, \;\; M_{W_R} =\frac{g}{2}v_{R}, \;\;M_{Z_L} = \frac{g}{2} v_{L} \sqrt{1+ \frac{g^2_1}{g^2+g^2_1}}, \;\; M_{Z_R} = \frac{v_{R}}{2} \sqrt{(g^2+g^2_1)} $$
\begin{figure}[htb]
\centering
\includegraphics[scale=0.95]{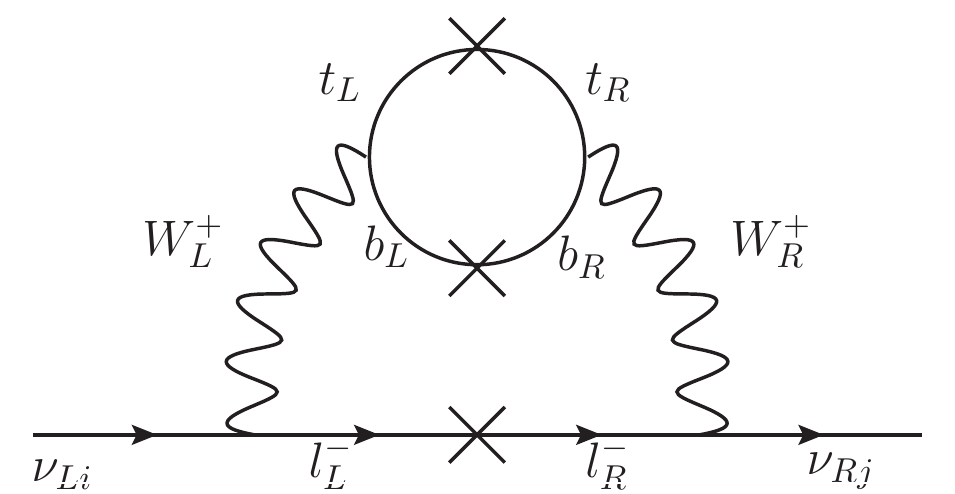}
\caption{Two-loop contribution to Dirac neutrino mass}
\label{numass2}
\end{figure}

The neutrinos however, remain massless at tree level due to the absence of the corresponding heavy neutral fermions. However, $\nu_L$ can acquire a Dirac mass through mixing with $N_R$ at two loop level, as seen from figure \ref{numass2}. The possibility of generating Dirac neutrino mass through this diagram was first proposed by \cite{balamoha} and the corresponding two loop Dirac neutrino mass can be estimated to be approximately
\begin{equation}
M_{LR} \approx \frac{\alpha m_{l^-}}{4 \pi \sin^2{\theta_W}} \theta_{L-R} I
\end{equation}
where $I$ is the loop integration factor (of the order $1-10$) and $\theta_{L-R}$ is the one loop mixing between $W_L, W_R$ given by 
\begin{equation}
\theta_{L-R} \approx \frac{\alpha}{4 \pi \sin^2{\theta_W}} \frac{m_b m_t}{M^2_{W_R}}
\end{equation}
Using $\alpha = 1/137, \sin^2{\theta_W} \approx 0.23, m_b \approx 4.2\; \text{GeV}, m_t \approx 174 \; \text{GeV}, M_{W_R} \approx 3\; \text{TeV}$, we find $\theta_{L-R} \approx 2 \times 10^{-7}$. Using this in the expression for Dirac mass we get 
\begin{equation}
M_{LR} \approx (1-10) \times 5.2 \times 10^{-10} m_{l^-}
\end{equation}
which, for $m_{l^-} = m_e \approx 0.5 \; \text{MeV}$ becomes $M_{LR} \approx (1-10) \times 2.6 \times 10^{-4} \; \text{eV}$. On the other hand, for $m_{l^-} = m_{\tau} \approx 1.77 \; \text{GeV}$, the Dirac mass becomes $M_{LR} \approx (1-10) \times 0.92 \; \text{eV}$. Instead of using the approximate formula for this two loop Dirac mass from \cite{balamoha, babuhe} for qualitative understanding, we can also derive the exact formula given by 
\begin{align}
M_{LR} &= \frac{\alpha m_{l^-}}{4 \pi \sin^2{\theta_W}} \frac{\sin{2 \theta_{L-R}}}{2} \left(f(x_{l,W_R}) - f(x_{l,WL})\right) \\
\sin{2\theta_{L-R}} &=\frac{2W_{LR}}{\sqrt{\left(M^2_{W_R}-M^2_{W_L}\right)^2 + 4W^2_{LR}}} \nonumber \\ 
W_{LR} &= \frac{4\pi \alpha}{\sin^2 \theta_{W}}\sum_{u,d}m_u m_d V_{u,d}V^*_{u,d}f(x_{u,d}); \quad x_{i,j} = \frac{m^2_i}{m^2_j}\nonumber \\
f(x_{i,j}) &= \frac{1}{16\pi^2}\left[\frac{x_{i,j}\ln (x_{i,j}) + 1 - x_{i,j}}{1-x_{i,j}}+\ln \left(\frac{\mu^2}{m^2_{j}}\right)\right] \nonumber
\end{align}
Although such two loop diagrams can generate neutrino masses of the correct order, it fails to generate the observed leptonic mixing. This is due to the proportionality between light neutrino mass and charged lepton mass $M_{LR} \propto m_{l^-}$. Since the leptonic mixing matrix is $U=U^{\dagger}_{lL} U_{\nu L}$, this proportionality relation results in a diagonal leptonic mixing matrix, ruled out by the data from neutrino oscillation experiments suggesting large leptonic mixing. This motivates us to go beyond this minimal model in order to generate tiny Dirac neutrino mass along with correct leptonic mixing.

\section{Higgs Doublet Extension of Minimal Model}
\label{sec1b}
Since the minimal LRSM with universal seesaw for charged fermions and radiative Dirac neutrino mass for light neutrinos can not give rise to correct leptonic mixing as discussed above, here we consider a minimal extension of this model. As shown recently \cite{db1, db3}, the minimal model can be extended by heavy neutral fermions having non-trivial transformations under additional discrete symmetries higher than $Z_2$ in such a way that a tiny Dirac neutrino mass can be generated at one-loop level. Here we consider a simpler scenario without any such discrete symmetries and additional heavy neutral fermions, as proposed by \cite{ma1989} many years back. We will incorporate an additional discrete $Z_2$ symmetry only when we extend our model to include a stable dark matter candidate through \textit{scotogenic} fashion \cite{ma06}.

The particle content of this model is the extension of the one shown in table \ref{tab:data1} and \ref{tab:data2} by another pair of scalar doublets.  We denote the four scalar doublets in this model as $H_{1L, 2L, 1R, 2R}$.  Similar to the minimal model, here also the neutral components of these scalars acquire non-zero vev's in order to break the gauge symmetry of LRSM all the way down to the $SU(3)_c \times U(1)_Q$ leading to heavy vector bosons $W_{L,R}, Z_{L,R}$. The Lagrangian for fermions can be written as
\begin{align}
\mathcal{L} & \supset Y_{Ui} (\overline{q_L} H_{iL} U_L+\overline{q_R} H_{iR} U_R) + Y_{Di} (\overline{q_L} H^{\dagger}_{iL} D_L+\overline{q_R} H^{\dagger}_{iR} D_R) +M_U \overline{U_L} U_R+ M_D \overline{D_L} D_R\nonumber \\
& +Y_{Ei} (\overline{\ell_L} H^{\dagger}_{iL} E_L+\overline{\ell_R} H^{\dagger}_{iR} E_R) +M_E \overline{E_L} E_R+\text{h.c.}
\label{fermionL}
\end{align}

The scalar Lagrangian can be written as
\begin{equation}
V = V_{L} + V_{R} + V_{LR}
\end{equation}
\begin{equation}
V_{L} = \mu^2_{Lij} (H^{\dagger}_{iL} H_{jL}) +\lambda_{Lijkl} (H^{\dagger}_{iL} H_{jL})(H^{\dagger}_{kL} H_{lL}) 
\end{equation}
\begin{equation}
V_{R} = \mu^2_{Rij} (H^{\dagger}_{iR} H_{jR}) +\lambda_{Rijkl} (H^{\dagger}_{iR} H_{jR})(H^{\dagger}_{kR} H_{lR}) 
\end{equation}
\begin{equation}
V_{LR}=\lambda_{LRijkl}  (H^{\dagger}_{iL} H_{jL})(H^{\dagger}_{kR} h_{lR}) 
\end{equation}

Denoting the vacuum expectation value (vev) acquired by the neutral components of the scalar doublets as $\langle H^0_{iL} \rangle = v_{iL}/\sqrt{2}, \langle H^0_{iR} \rangle = v_{iR}/\sqrt{2}$, one can write down the gauge symmetry breaking chain as 
$$SU(2)_L \times SU(2)_R \times U(1)_{B-L} \quad \underrightarrow{\langle
H_{iR} \rangle} \quad SU(2)_L\times U(1)_Y  \quad \underrightarrow{\langle H_{iL} \rangle} \quad U(1)_{Q}$$
where the $SU(3)_c$ is not mentioned as it remains unbroken throughout. At the end of the symmetry breaking, different components of the above multiplets acquire electromagnetic charged $Q$ according to the following relation 
\begin{align}
Q=T_{3L}+T_{3R}+\frac{B-L}{2} 
\end{align}
As a result of this symmetry breaking, two charged and two neutral bosons acquire masses. The charged boson mass matrix squared in $W^{\pm}_L, W^{\pm}_R$ basis is given by
\begin{equation}
M^2_{\pm} = \frac{1}{4}\begin{pmatrix} g^2_L (v^2_{1L}+v^2_{2L}) & 0 \\ 0 & g^2_R (v^2_{1R}+v^2_{2R}) \end{pmatrix}
\end{equation}
The corresponding neutral gauge boson mass matrix squared in the $(W_{L3}, W_{R3}, B)$ basis is
\begin{equation}
M^2_{0} = \frac{1}{4}\begin{pmatrix} g^2_L (v^2_{1L}+v^2_{2L}) & 0 & -g_1 g_L (v^2_{1L}+v^2_{2L}) \\ 0 & g^2_R (v^2_{1R}+v^2_{2R}) & -g_1 g_R (v^2_{1R}+v^2_{2R})\\  -g_1 g_L (v^2_{1L}+v^2_{2L}) & -g_1 g_R (v^2_{1R}+v^2_{2R}) & g^2_1(v^2_{1L}+v^2_{2L}+v^2_{1R}+v^2_{2R})\end{pmatrix}
\end{equation}
In the left-right symmetric limit $g_L = g_R = g$ and assuming $v_{1L, 2L} \ll v_{1R, 2R}$, the vector boson masses can be written as 
$$ M_{W_L} = \frac{g}{2}\sqrt{v^2_{1L}+v^2_{2L}}, \;\;\; M_{W_R} =\frac{g}{2}\sqrt{v^2_{1R} + v^2_{2R}} $$
$$M_{Z_L} = \frac{g}{2} \sqrt{v^2_{1L}+v^2_{2L}} \sqrt{1+ \frac{g^2_1}{g^2+g^2_1}}, \;\;\; M_{Z_R} = \frac{1}{2} \sqrt{(g^2+g^2_1)(v^2_{1R} + v^2_{2R})} $$

\begin{figure}[!h]
\centering
\epsfig{file=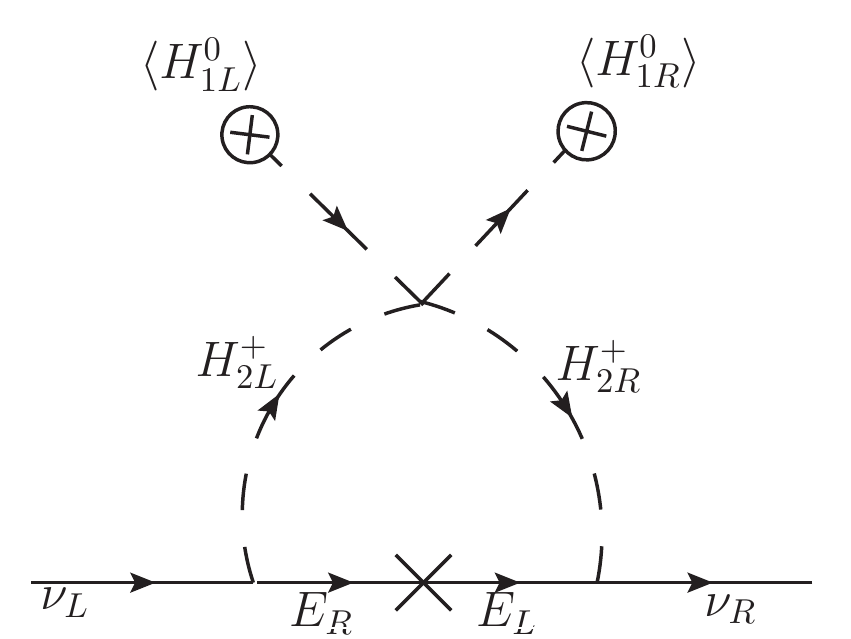,width=0.50\textwidth,clip=}
\caption{One-loop Dirac neutrino mass for the particle content shown in table \ref{tab:data1}}
\label{fig1}
\end{figure}

The resulting charged fermion masses are given in a similar way as the minimal model as
$$m_u = \sum_{i,j=1,2} Y_{Ui} \frac{1}{M_U} Y^T_{Uj} v_{iL} v_{jR}, \;\; m_d = \sum_{i,j=1,2} Y_{Di} \frac{1}{M_D} Y^T_{Dj} v_{iL} v_{jR}, \;\; m_e = \sum_{i,j=1,2} Y_{Ei} \frac{1}{M_E} Y^T_{Ej} v_{iL} v_{jR} $$
Since there are no heavy neutral fermions for neutrinos to couple to, they remain massless at tree level. However, at one loop level they can acquire a non-zero mass through the mass diagram shown in figure \ref{fig1}. This neutrino mass diagram is somewhat similar to the one in Zee model \cite{Zee} of radiative neutrino mass except the way the scalar vertex is completed at the top. Here it is a quartic scalar vertex while in the Zee model, it is a trilinear one. The one loop mass corresponding to the diagram in figure \ref{fig1} can be estimated as 
\begin{equation}
(m_{\nu})_{ij} =\frac{\sin{\theta_1} \cos{\theta_1}}{32 \pi^2} \sum_k (Y_{E2})_{ik}(Y_{E2})_{kj} (M_{E})_k \left ( \frac{m^2_{\xi^+_1}}{m^2_{\xi^+_1}-(M_E)^2_k} \text{ln} \frac{m^2_{\xi^+_1}}{(M_E)^2_{k}}-\frac{m^2_{\xi^+_2}}{m^2_{\xi^+_2}-(M_E)^2_{k}} \text{ln} \frac{m^2_{\xi^+_2}}{(M_E)^2_{k}} \right)
\label{numassR}
\end{equation}
where $\theta_1$ is the mixing angle between $H^+_{2L}, H^+_{2R}$ given by
$$\tan{2 \theta_1} = \frac{\lambda_{LR} v_{1L} v_{1R}}{m^2_{\xi^+_2}-m^2_{\xi^+_1}} $$
and $\xi^+_{1,2}$ corresponds to the mass eigenstates of $H^+_{2L}, H^+_{2R}$. Assuming $m^2_{\xi^+_1}, m^2_{\xi^+_2} \ll M_E$ we can write down the most dominant contribution to light neutrino mass as 
\begin{equation}
(m_{\nu})_{ij} \approx \frac{\sin{\theta_1} \cos{\theta_1}}{64 \pi^2} \sum_k (Y_{E2})_{ik}(Y_{E2})_{kj} \left ( \frac{m^2_{\xi^+_1}+m^2_{\xi^+_2}}{(M_E)_k} \text{ln} \frac{m^2_{\xi^+_2}}{m^2_{\xi^+_1}} \right)
\label{numassR1}
\end{equation}
For small mixing angle $\theta_1$, all Yukawa couplings $Y_{E1} \approx Y_{E2}$ and vev's of similar magnitudes $v_{1L} \approx v_{2L}, v_{1R} \approx v_{2R}$ and ignoring the generation indices, we can write down the neutrino mass in terms of charged lepton mass as
\begin{equation}
m_{\nu} \approx  m_e \frac{\lambda_{LR}(m^2_{\xi^+_1}+m^2_{\xi^+_2})}{128 \pi^2 (m^2_{\xi^+_2}-m^2_{\xi^+_1})} \text{ln} \frac{m^2_{\xi^+_2}}{m^2_{\xi^+_1}}
\end{equation}
which will require substantial fine-tuning in quartic coupling $\lambda_{LR} \sim 10^{-7}$ in order to generate the hierarchy between tau lepton and neutrinos. This can however be improved by going invoking inequalities of Yukawa couplings and the vev's. The interesting feature of such a scenario is that the same vector like fermions and Yukawa couplings responsible for charged lepton masses also generate tiny neutrino masses without the need of any additional field content.

\section{Scotogenic LRSM}
\label{sec2}

\begin{figure}[!h]
\centering
\epsfig{file=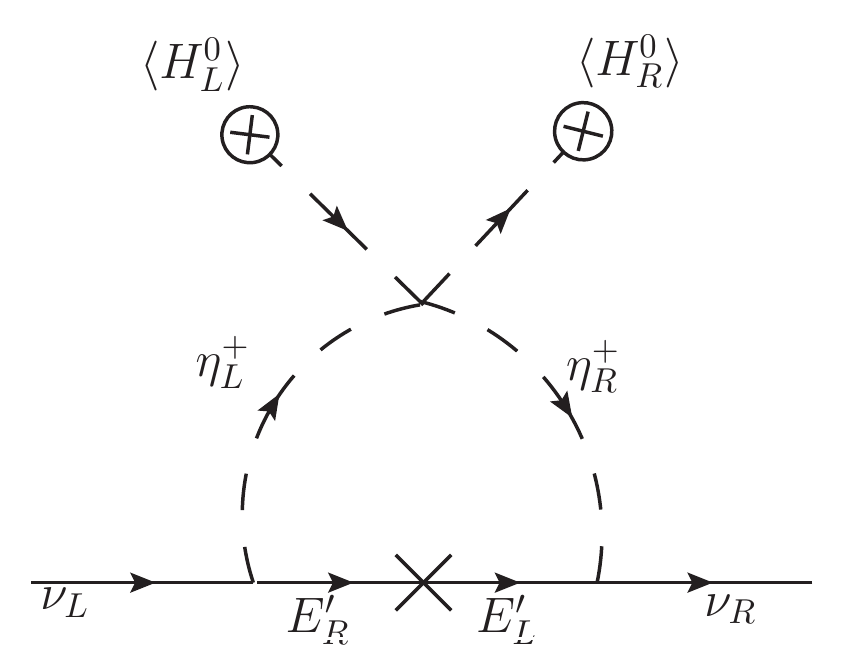,width=0.50\textwidth,clip=}
\caption{One-loop Dirac neutrino mass for the scotogenic extension of the particle content shown in table \ref{tab:data1}}
\label{fig2}
\end{figure}

The minimal LRSM discussed above with one loop Dirac neutrino mass does not have a dark matter candidate. One can however, extend the above model in a minimal manner to incorporate dark matter. One such minimal way is to include additional scalar or fermionic multiplets in the spirit of minimal dark matter scenario \cite{Cirelli:2005uq,Garcia-Cely:2015dda,Cirelli:2015bda}. Such minimal dark matter scenario in LRSM has been studied recently by the authors of \cite{Heeck:2015qra,Garcia-Cely:2015quu}. In these models, the dark matter candidate is stabilised either by a $Z_2 = (-1)^{B-L}$ subgroup of the $U(1)_{B-L}$ gauge symmetry or due to an accidental symmetry at the renormalisable level due to the absence of any renormalisable operator leading to dark matter decay \cite{db3}. Another possibility to incorporate dark matter into the above models is to generate the one loop Dirac neutrino mass in a scotogenic fashion \cite{ma1} where the same scalar multiplet is responsible for one-loop neutrino mass as well as giving rise to a stable dark matter candidate. The minimal model discussed above has to be extended by an additional $Z_2$ symmetry and at least two copies of vector like charged fermions $E^{\prime}_{L,R}$ which are odd under this $Z_2$ symmetry. The second Higgs doublet in both left and right handed sectors are also assigned $Z_2$ odd charges in order to make them stable, if lighter than $E^{\prime}_{L,R}$. We denote the second copy of these scalar doublets as $\eta_{L,R}$ just to distinguish them from the usual scalar doublets $H_{L,R}$ taking part in the symmetry breaking. It should be noted that, adding two (three) copies of neutral vector like fermions $N_{L,R}$ at this stage can not ensure a pure Dirac type light neutrino as $Z_2$ symmetry is not enough to keep the Majorana mass terms of $N_L N_L$ or $N_R N_R$ type away from the Lagrangian. As stressed earlier, one requires higher symmetries than $Z_2$ to realise a pure Dirac type one-loop neutrino mass, in models with additional heavy neutral fermions. With this minimal modifications, the gauge boson and charged fermion masses arises in a similar way as in the minimal model discussed in section \ref{sec1}. The neutrinos acquire a Dirac mass term at one loop level through the diagram shown in figure \ref{fig2} which can be estimated in a way similar to the discussion above.

The lightest $Z_2$ odd particle is naturally stable in this model and hence can be a dark matter candidate. Since dark matter has to be neutral, one has to make sure that the vector like charged leptons $E^{\prime}$ are heavier than the neutral component of at least one of the $Z_2$ odd scalar doublets $\eta_{L,R}$. If $m_{\eta^0_L} < m_{\eta^0_R}$, then the dark matter candidate is similar to the inert scalar doublet model studied extensively in the literature \cite{ma06,Barbieri:2006dq,Majumdar:2006nt,LopezHonorez:2006gr,ictp,borahcline, honorez1,DBAD14}.

\section{Cosmological Implications}
\label{sec2b}
The minimal models discussed in this work both with and without dark matter can have very interesting implications for cosmology. Firstly because of the Dirac nature of the light neutrinos and secondly from dark matter point of view. The right handed components of Dirac neutrinos extra relativistic degrees of freedom which is strongly constrained from the Planck data \cite{Planck15}. Thus us due to the fact that such extra relativistic degrees of freedom may affect the big bang nucleosynthesis (BBN) predictions as well as cause changes in the cosmic microwave background (CMB) spectrum, which are very accurately measured. The $95\%$ limit on the effective number of relativistic degrees of freedom is \cite{Planck15}
\begin{equation}
N_{\text{eff}} = 3.15 \pm 0.23 \;\; (\text{Planck TT+lowP+BAO})
\label{neffbound1}
\end{equation}
which is consistent with the standard value $N_{\text{eff}} = 3.046$. Here the keywords in parenthesis refer to different constraints imposed to obtain the bound, the details of which can be found in \cite{Planck15}. Since the Dirac nature of light neutrinos double the relativistic degrees of freedom in comparison to the scenario with Majorana light neutrinos, this may well be in conflict with the Planck data. However, such scenario can still be consistent with these bounds if the right handed neutrinos decouple much earlier compared to the standard left handed neutrinos. This can be ensured due to the fact that left and right handed neutrinos have different gauge interactions in LRSM and the strength of such interactions decides the decoupling temperature of the respective species. If $T^D_{\nu_R} > T^D_{\nu_L}$, then the $\nu_R$ contribution to $N_{\text{eff}}$ gets diluted due to the decrease in effective relativistic degrees of freedom $g_*$ as the Universe cools down from $T^D_{\nu_R}$ to $T^D_{\nu_L}$ \cite{nuRdecoupling}:
$$ N_{\text{eff}} \approx 3 + 3 \left (\frac{g_*(T^D_{\nu_L})}{g_*(T^D_{\nu_R})} \right )^{\frac{4}{3}} $$
Here $g_*(T^D_{\nu_L}) = 10.75$ for the relativistic degrees of freedom in SM at $T=T^D_{\nu_L} \approx 1$ MeV. Using the conservative value $N_{\text{eff}} = 3.046$, it is straightforward to find that the light right handed neutrinos can be consistent with this if $g_*(T^D_{\nu_R}) > 246.71$ which is more than the relativistic degrees of freedom of all the SM particles $g_* (\text{SM}) = 106.75$ at temperatures above the electroweak scale. The decoupling temperature of right handed neutrinos can be calculated by following the same procedure for left handed neutrinos and replacing the $W_L$ mass with $W_R$. In terms of $T^D_{\nu_L}$, it can be written as
$$ T^D_{\nu_R} \approx (g_*(T^D_{\nu_R})^{1/6} \left ( \frac{M_{W_R}}{M_{W_L}} \right )^{4/3}  T^D_{\nu_L} $$
Demanding $T^D_{\nu_R} \gg 300$ GeV and taking $T^D_{\nu_L} \approx 1$ MeV, we can arrive at the bound on $M_{W_R}$ as 
\begin{equation}
M_{W_R} \gg 522 \; \text{TeV}
\end{equation}
which is way outside the reach of present or near future experimental reach. Taking the maximum possible value from the bound given in \eqref{neffbound1} that is $N_{\text{eff}} = 3.38$, we find $g_*(T^D_{\nu_R}) > 50.63$ which corresponds to a decoupling temperature of around $0.3$ GeV. This corresponds to a bound on the $W_R$ mass 
\begin{equation}
M_{W_R} > 3.548 \; \text{TeV}
\end{equation}
which lies within the reach of ongoing collider experiment. A slightly more strict bound on the number of relativistic degrees of freedom imposing different sets of constraints is found to be \cite{Planck15}
\begin{equation}
N_{\text{eff}} = 3.04 \pm 0.18 \;\; (\text{Planck TT, TE, EE+lowP+BAO})
\label{neffbound2}
\end{equation}
Taking the maximum possible value of effective relativistic degrees of freedom $N_{\text{eff}} = 3.22$, we find $g_*(T^D_{\nu_R}) > 76.28$ which corresponds to a decoupling temperature of around 1 GeV. This implies that the right handed neutrinos should decouple before the QCD phase transition temperature of around 200-400 MeV. Demanding $T^D_{\nu_R} > 1 $ GeV, this gives rise to a bound on $W_R$ mass as
\begin{equation}
M_{W_R} > 8.387 \; \text{TeV}
\end{equation}
which can be accessible at high energy future colliders.

The other cosmological implication is dark matter, which is incorporated in the scotogenic version of the model discussed above. According to the latest Planck data, around $26\%$ of the present Universe's energy density being made up of dark matter \cite{Planck15}. Their estimate can also be expressed in terms of density parameter $\Omega$ as
\begin{equation}
\Omega_{\text{DM}} h^2 = 0.1187 \pm 0.0017
\label{dm_relic}
\end{equation}
where $h = \text{(Hubble Parameter)}/100$ is a parameter of order unity. In the scotogenic LRSM studied here, the lighter of the neutral (pseudo) scalars in $\eta^0_{L,R}$ is stabilised by the unbroken $Z_2$ symmetry of the model, and hence can be a stable dark matter candidate. While the phenomenology of $\eta^0_L$ dark matter is same as the inert doublet model (IDM) \cite{ma06,Barbieri:2006dq,Majumdar:2006nt,LopezHonorez:2006gr,ictp,borahcline, honorez1,DBAD14}, the right handed scalar doublet dark matter $\eta^0_R$ was studied recently by \cite{db3, Garcia-Cely:2015quu}, considering only the gauge annihilation diagrams. In the present model however, there are additional diagrams corresponding to dark matter annihilations into charged fermions through the exchange of vector like charged leptons $E^{\prime}_{L,R}$. Such t-channel diagrams, though remain suppressed compared to the gauge annihilation diagrams, can play a non-trivial role for those region of dark matter masses where the latter remains sub-dominant. This will also allow us to relate the parameter space satisfying neutrino data and dark matter relic abundance, as the same couplings and masses determine them. Although here, we do not perform a general scan of parameters satisfying neutrino and dark matter data, we calculate the relic abundance of both the dark matter candidates $\eta^0_{L,R}$ including the new annihilation channel through vector like leptons and also incorporating the bound on $W_R$ mass from the bound on $N_{\text{eff}}$ discussed above.

The dark matter candidate in our model is similar to a typical weakly interacting massive particle (WIMP) candidate which was in thermal equilibrium in the early Universe but decouple from the plasma subsequently when the rate of interaction falls below the expansion rate of the Universe. Its relic abundance can be calculated by solving the relevant Boltzmann equation 
\begin{equation}
\frac{dn_{\chi}}{dt}+3Hn_{\chi} = -\langle \sigma v \rangle (n^2_{\chi} -(n^{\text{eqb}}_{\chi})^2)
\label{BE1}
\end{equation}
where $n_{\chi}$ is the dark matter number density and $n^{eqb}_{\chi}$ is the corresponding equilibrium number density. $H$ is the Hubble expansion rate of the Universe and $ \langle \sigma v \rangle $ is the thermally averaged annihilation cross section of the dark matter particle $\chi$. For most practical purposes, it is sufficient to use the approximate analytical solution of the above Boltzmann equation given by \cite{Kolb:1990vq, kolbnturner}
\begin{equation}
\Omega_{\chi} h^2 \approx \frac{1.04 \times 10^9 x_F}{M_{Pl} \sqrt{g_*} (a+3b/x_F)}
\end{equation}
where $x_F = m_{\chi}/T_F$, $m_{\chi}$ is the mass of Dark Matter particle, $T_F$ is the freeze-out temperature, $g_*$ is the number of relativistic degrees of freedom at the time of freeze-out and $M_{Pl} \approx 10^{19}$ GeV is the Planck mass. Here $a, b$ are the coefficients of partial wave expansion $ \langle \sigma v \rangle = a +b v^2$. The freeze-out temperature or $x_F$ can be calculated iteratively from the following relation 
\begin{equation}
x_F = \ln \frac{0.038gM_{\text{Pl}}m_{\chi}<\sigma v>}{g_*^{1/2}x_F^{1/2}}
\label{xf}
\end{equation}
which follows from equating the interaction rate with the Hubble expansion rate of the Universe. The thermal averaged annihilation cross section $\langle \sigma v \rangle$ is given by \cite{Gondolo:1990dk}
\begin{equation}
\langle \sigma v \rangle = \frac{1}{8m^4_{\chi}T K^2_2(m_{\chi}/T)} \int^{\infty}_{4m^2_{\chi}}\sigma (s-4m^2_{\chi})\surd{s}K_1(\surd{s}/T) ds
\label{eq:sigmav}
\end{equation}
where $K_i$'s are modified Bessel functions of order $i$. However, if one of the components of the dark matter multiplet has a mass close to the mass of the lightest neutral component or the DM, then those next to lightest component can be thermally accessible during DM freeze-out. In such a situation, the DM particle can undergo coannihilations with these heavier components which can significantly affect the final relic abundance of DM. This type of coannihilation effects on dark matter relic abundance were studied by several authors in \cite{Griest:1990kh, coann_others}. Here we follow the analysis of \cite{Griest:1990kh} to calculate the effective annihilation cross section in such a case. The effective cross section can given as 
\begin{align}
\sigma_{\text{eff}} &= \sum_{i,j}^{N}\langle \sigma_{ij} v\rangle r_ir_j \nonumber \\
&= \sum_{i,j}^{N}\langle \sigma_{ij}v\rangle \frac{g_ig_j}{g^2_{eff}}(1+\Delta_i)^{3/2}(1+\Delta_j)^{3/2}e^{\big(-x_F(\Delta_i + \Delta_j)\big)} \nonumber \\
\end{align}
where, $x_F = \frac{m_{DM}}{T_F}$ and $\Delta_i = \frac{m_i-m_{DM}}{m_{DM}}$  and 
\begin{align}
g_{\text{eff}} &= \sum_{i=1}^{N}g_i(1+\Delta_i)^{3/2}e^{-x_F\Delta_i}
\end{align}
The thermally averaged cross section can be written as
\begin{align}
\langle \sigma_{ij} v \rangle &= \frac{x_F}{8m^2_im^2_jm_{DM}K_2((m_i/m_{DM})x_F)K_2((m_j/m_{DM})x_F)} \times \nonumber \\
& \int^{\infty}_{(m_i+m_j)^2}ds \sigma_{ij}(s-2(m_i^2+m_j^2)) \sqrt{s}K_1(\sqrt{s}x_F/m_{DM}) \nonumber \\
\label{eq:thcs}
\end{align}
Since the effective coupling $g_{eff}$ decreases exponentially with increasing mass splitting between DM and heavier components, one can ignore such coannihilation effects for scenarios with large mass splitting.
\begin{figure}[!h]
\centering
\begin{tabular}{cc}
\epsfig{file=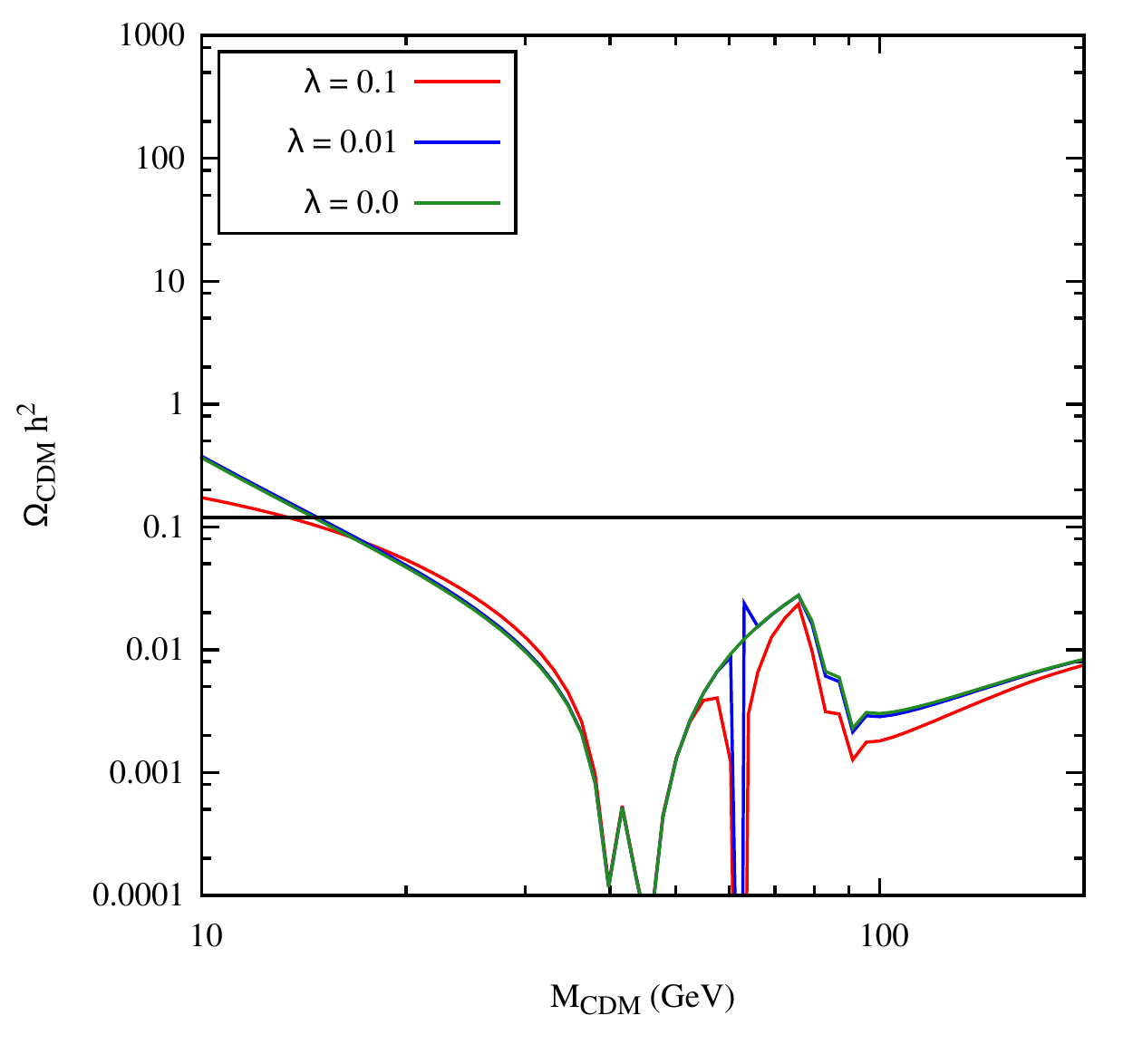,width=0.50\textwidth,clip=}
\epsfig{file=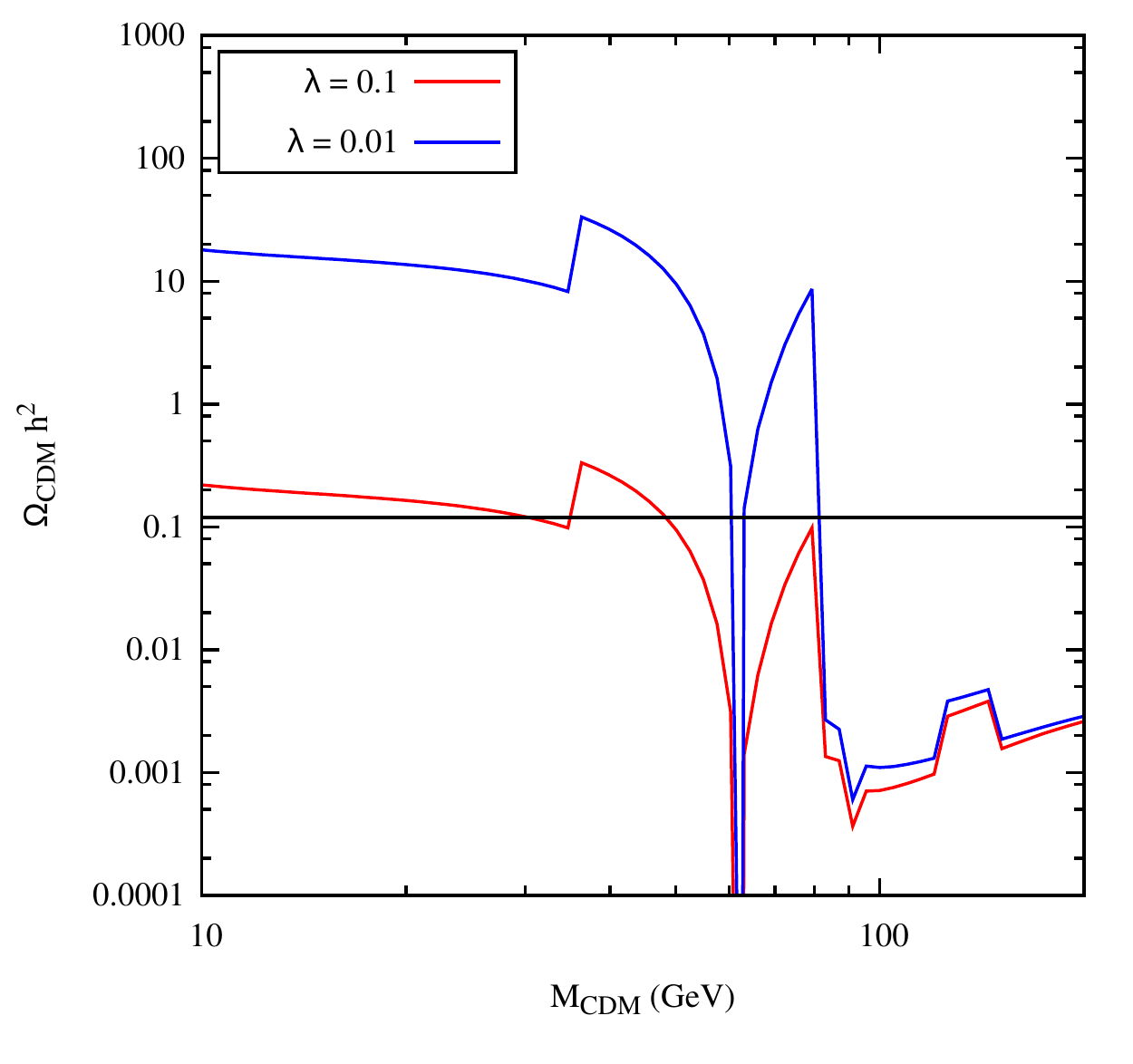,width=0.50\textwidth,clip=}
\end{tabular}
\caption{Relic abundance of $\eta^0_L$ dark matter as a function of its mass. The mass splitting of $\eta^0_R$ with charged and pseudoscalar components of $\eta_L$ doublet is fixed at 1 GeV, 50 GeV in the left and right panel plots respectively.}
\label{fig2b}
\end{figure}

\begin{figure}[!h]
\centering
\epsfig{file=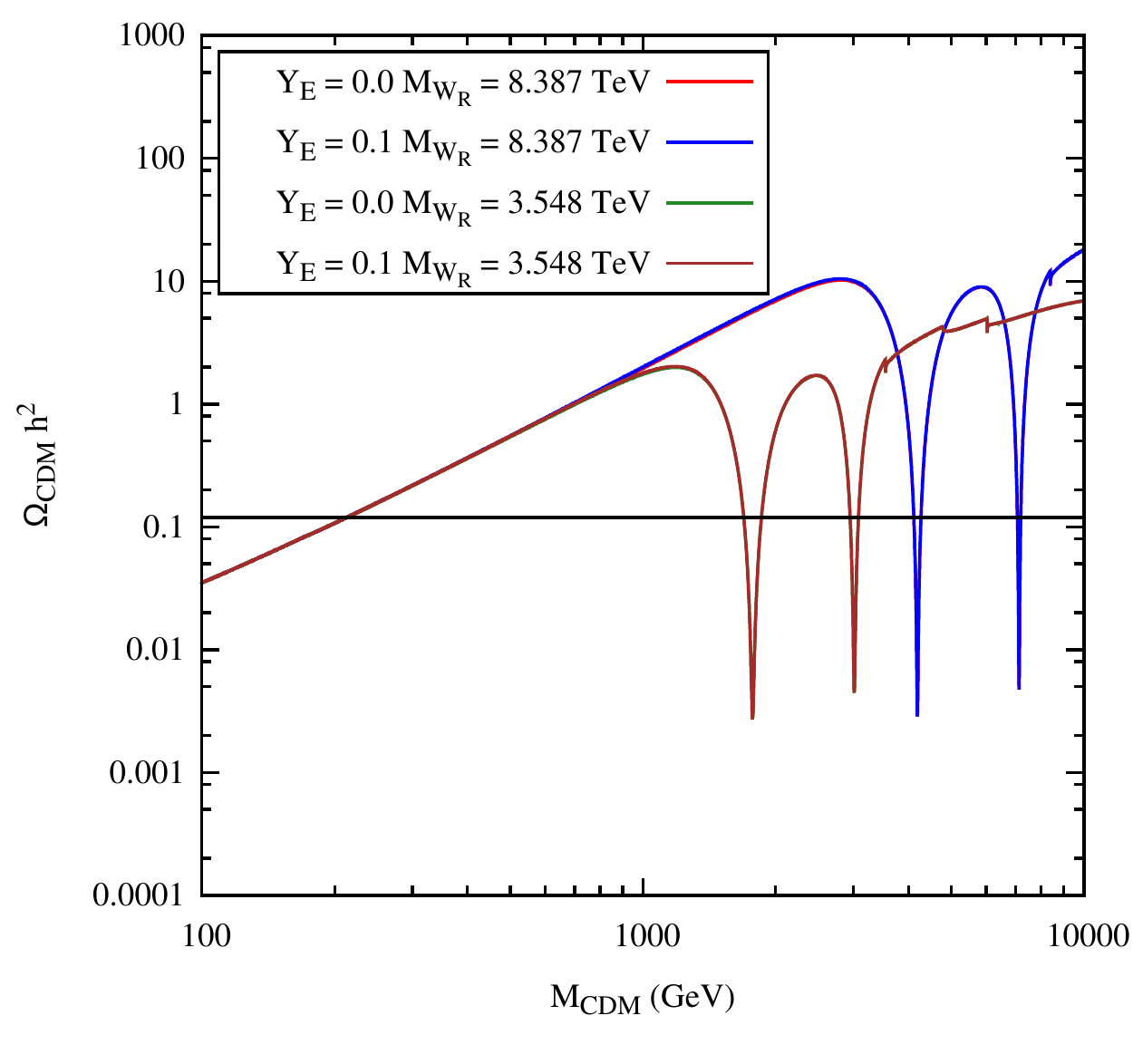,width=0.90\textwidth,clip=}
\caption{Relic abundance of $\eta^0_R$ dark matter as a function of its mass. The mass splitting of $\eta^0_R$ with charged and pseudoscalar components of $\eta_R$ doublet is fixed at 1 GeV. The vector like charged lepton masses are taken to be 500 GeV. }
\label{fig2a}
\end{figure}
We calculate the relic abundance of both $\eta^0_L$ and $\eta^0_R$ dark matter as a function of their masses. Though only the lighter of them is stabilised by the $Z_2$ symmetry, we do not assume any particular mass relations between them and calculate their relic abundance individually. The resulting parameter space then gives an idea about the masses of $\eta^0_L$ and $\eta^0_R$ that can satisfy the relic abundance criteria. In case of $\eta^0_L$ dark matter, there are two mass ranges of dark matter where the relic abundance criteria is satisfied, as studied extensively in the context of IDM. One is lthe ow mass regime $m_{\eta^0_L}=m_{DM} \leq M_W$ and the other is the high mass regime $m_{\eta^0_L}=m_{DM} > 500$ GeV. The main annihilation or coannihilation channels affecting the relic abundance of $\eta^0_L$ dark matter are the ones mediated by the electroweak $W_L, Z_L$ bosons or the standard model Higgs boson. The additional new physics introduced in this work for example, the vector like lepton's contribution to the relic abundance usually remain suppressed compared to the above mentioned channels. Also, the right handed gauge sector is unlikely to affect $\eta^0_L$ abundance due to tiny left-right mixing at one-loop $\theta_{L-R} \approx 10^{-7}$ for TeV scale $W_R$ mass discussed in the context of the minimal model in section \ref{sec1}. We show the relic abundance of $\eta^0_L$ dark matter in the low mass regime as a function of its mass in the plot shown in figure \ref{fig2b}. In the left panel of figure \ref{fig2b}, the mass splitting between different components of $\eta_L$ doublet is assumed to be 1 GeV and the same for the right panel plot is fixed at 50 GeV. As expected, the smaller mass difference enhances the coannihilations, reducing the dark matter relic abundance. We also include the vector like charged lepton contribution by assuming their masses to be 500 GeV with order one Yukawa couplings. However, those t channel diagrams remain suppressed and do not affect the dark matter relic abundance significantly.

The relic abundance of $\eta^0_R$ dark matter on the other hand, is sensitive to the scale of left-right symmetry or the mass of $W_R$. Since the mass of $W_R$ is also constrained from the Planck bound on $N_{\text{eff}}$ discussed above, the scenario with $\eta^0_R$ can be constrained from the bounds on dark matter relic abundance as well as $N_{\text{eff}}$ simultaneously. It is interesting to note that the cosmology bound on the $W_R$ mass is stronger than other experimental bounds. While $K-\bar{K}$ mixing puts a constraint $M_{W_R} > 2.5$ TeV \cite{kkbar}, the dijet searches at ATLAS experiment of the LHC at 13 TeV centre of mass energy has excluded heavy $W_R$ boson masses below 2.9 TeV \cite{dijetATLAS2}. We first calculate the $\eta^0_R$ abundance for $M_{W_R}=3.548$ TeV, allowed by the latest direct search bounds at the LHC as well as the Planck bound on $N_{\text{eff}}$ given in \eqref{neffbound1} by considering the mass splitting between different components of $\eta_R$ doublet to be 1 GeV. Such a mass splitting keeps the coannihilation channels mediated by $W_R, Z_R$ efficient resulting in some region of $\eta^0_R$ mass where the relic abundance criteria can be satisfied. This can be seen from the plot shown in figure \ref{fig2a}. Similarly, we calculate the abundance of $\eta^0_R$ incorporating the more conservative bound on $W_R$ mass from the Planck constraints on $N_{\text{eff}}$ given in \eqref{neffbound2}. For both the values of $W_R$ mass, there is an allowed region of $\eta^0_R$ mass near 300 GeV where the relic abundance criteria is satisfied. There are two more allowed regions near $M_{W_R}/2$ and $M_{Z_R}$ respectively, as seen from the respective resonances in figure \ref{fig2a}. We have also included the annihilations mediated by vector like leptons $E^{\prime}$ considering the relevant Yukawa couplings to be $0.1$ and vector like lepton mass 500 GeV. However, as seen from the figure \ref{fig2a}, the inclusion of this additional t channel annihilation mediated by $E^{\prime}$ has negligible effect on the parameter space of $\eta^0_R$ dark matter.

\section{Flavour Physics Implications}
\label{sec3}

\begin{figure}[!h]
\centering
\epsfig{file=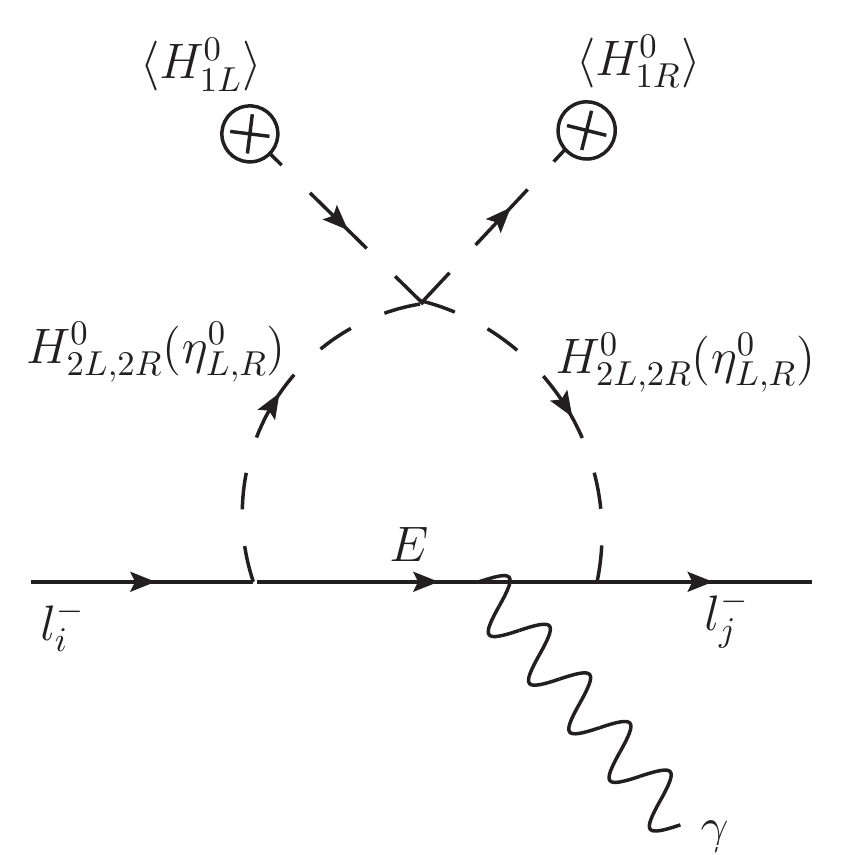,width=0.50\textwidth,clip=}
\caption{Lepton flavour violating decays of charged leptons at one loop.}
\label{fig3}
\end{figure}
\begin{figure}[!h]
\centering
\epsfig{file=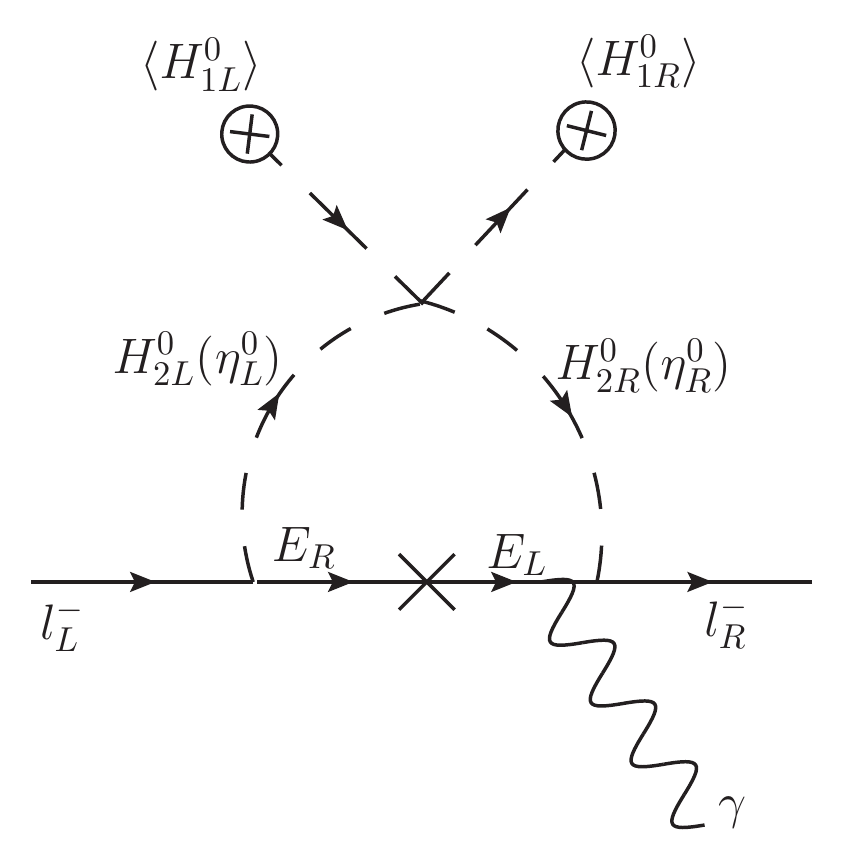,width=0.50\textwidth,clip=}
\caption{EDM of charged leptons}
\label{fig4}
\end{figure}
\begin{figure}[!h]
\centering
\epsfig{file=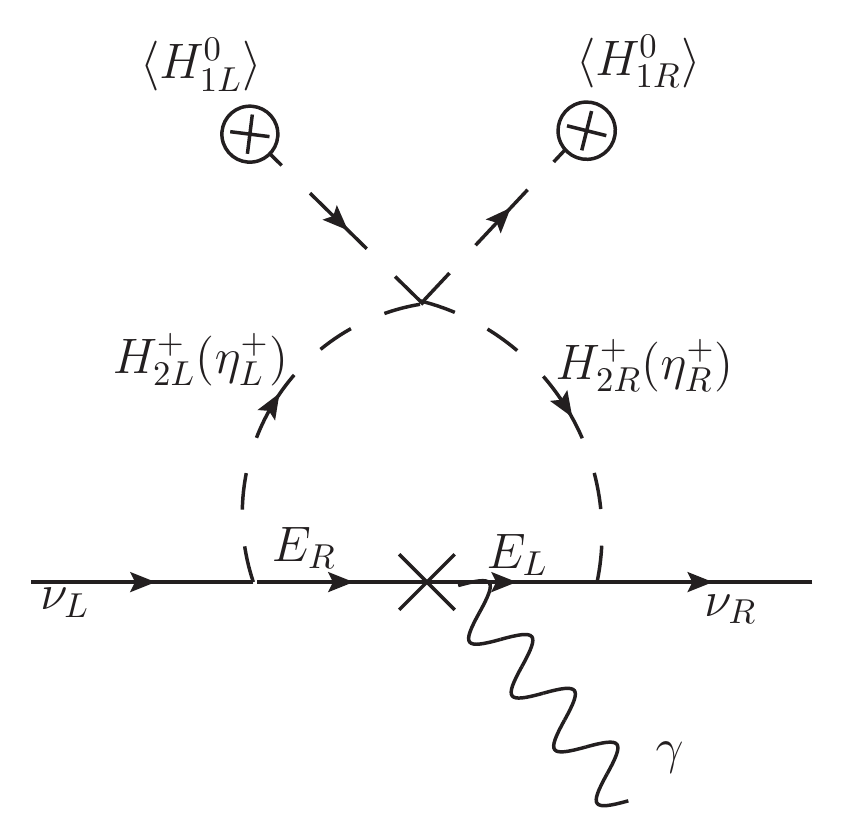,width=0.50\textwidth,clip=}
\caption{EDM of neutrinos at one loop.}
\label{fig5}
\end{figure}

\begin{figure}
\centering
\begin{tabular}{cc}
\epsfig{file=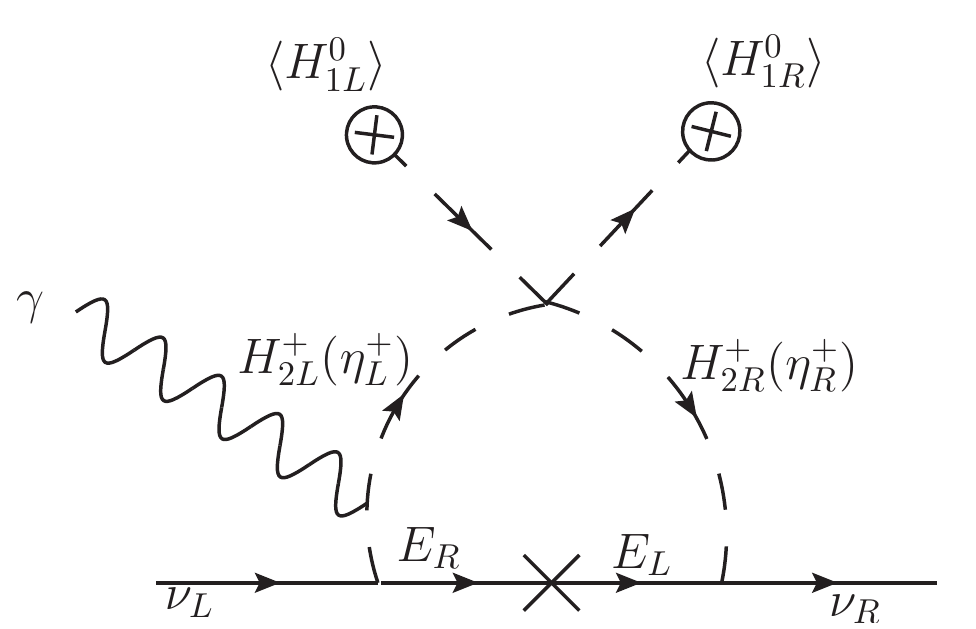,width=0.50\textwidth,clip=} &
\epsfig{file=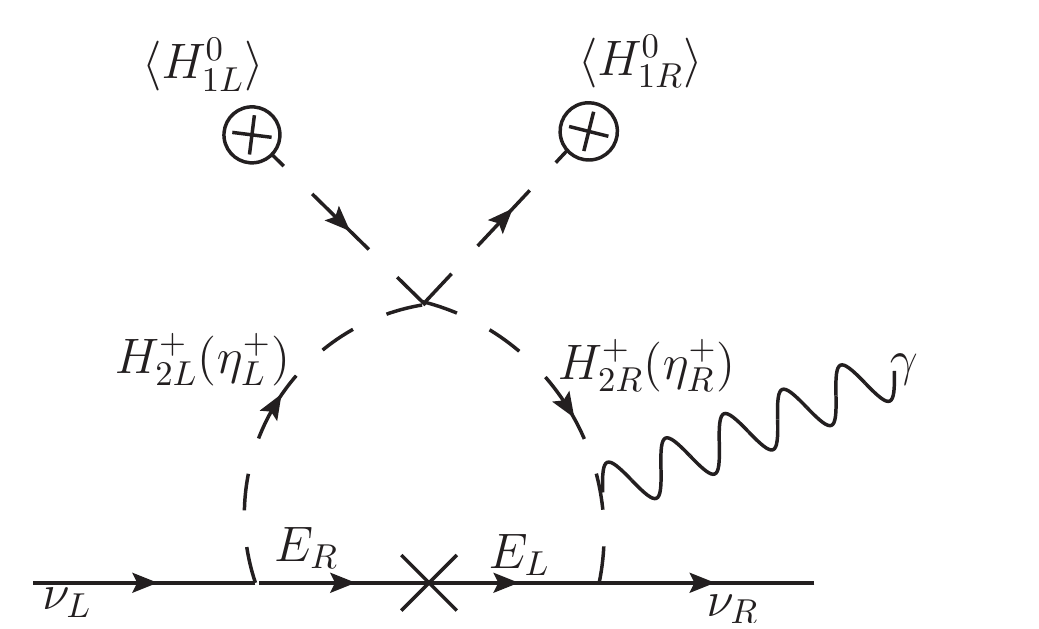,width=0.50\textwidth,clip=}
\end{tabular}
\caption{EDM of neutrinos at one loop.}
\label{fig5a}
\end{figure}

The other interesting phenomenology our model can have is in the flavour sector. One immediate consequence of these scenarios is the vanishing amplitude for neutrinoless double beta decay $(0\nu \beta \beta)$ which can be tested at ongoing or future experiments \cite{kamland, kamland2, GERDA, GERDA2, exo200, 0nbbexpt}. Any positive observation of $0\nu\beta \beta$ will indicate the Majorana nature of light neutrinos as required by the validity of the Schechter-Valle theorem \cite{schvalle}. However, the source of such observable $(0\nu \beta \beta)$ or other lepton number violating signatures at colliders may remain sub-dominant in their contributions to Majorana light neutrino masses keeping the light neutrino masses predominantly of Dirac type \cite{db3}.

Apart from the above testable prediction, the new fields introduced in the model can induce LFV decays like $\mu \rightarrow e \gamma$ \footnote{For a recent review on charged lepton flavour violation, please see \cite{PRlfv}} through one-loop diagrams with heavy charged vector like leptons and the second scalar doublets in loop. This is shown in figure \ref{fig3}. In the SM, such LFV decays also occur at loop level but heavily suppressed due to the smallness of neutrino masses, far beyond the current experimental sensitivity \cite{MEG16}. Therefore, any experimental observation of such rare decay processes will be a clear indication of BSM physics. We calculate the new physics contribution to $\Gamma (\mu \rightarrow e \gamma)$ and check for what values of new physics parameters, it can lie close to the latest bound from the MEG collaboration is $\text{BR}(\mu \rightarrow e \gamma) < 4.2 \times 10^{-13}$ at $90\%$ confidence level \cite{MEG16}. 

We adopt the general prescriptions given in \cite{LFV1} for the calculation of the LFV decay width. From the scalar potential written above, it is clear that the scalar fields can be rotated to their mass basis as 
\begin{align}
H^0_i &= O_{ij}\widetilde{H}^j = L_{ij}H^j_L + R_{ij}H^j_R.
\label{rotation-higgs}
\end{align}
Now, the Yukawa can be rewritten as
\begin{align}
\mathcal{L} \supset Y_{Ei}\sum_{i=1}^4(L^*_{ij} \overline{l}_LH^\dagger_j E_L + R^*_{ij} \overline{l}_RH^\dagger_j E_R ).
\end{align}
In order to derive processes contributing to LFV we need to use the above
interaction term and bring it in the following form
\begin{align}
-i \overline{l}_\beta (\sigma_L P_L + \sigma_R P_R)\sigma^{\mu \nu}l_\alpha F_{\mu \nu}
\end{align}
and to calculate the process $l_\alpha \rightarrow l_\beta \gamma$ we have the 
following expression
\begin{align}
\Gamma(l_\alpha \rightarrow l_\beta \gamma) &= \frac{\left(m^2_\alpha - m^2_\beta\right)^3}{4\pi m^3_\alpha}\left[|\sigma_L|^2 + |\sigma_R|^2\right]
\end{align}
where $\sigma_{L,R}$ is given as
\begin{align}
\sigma_L &= Q_F\left((m_\alpha O_{RR} + m_\beta O_{LL})g(t) + m_E O_{RL}h(t)\right) \nonumber \\
&+ Q_B\left((m_\alpha O_{RR} + m_\beta O_{LL})\overline{g}(t) + m_E O_{RL} \overline{h}(t)\right) \\
\sigma_R &= Q_F\left((m_\alpha O_{LL} + m_\beta O_{RR})g(t) + m_E O_{LR}h(t)\right) \nonumber \\
&+ Q_B\left((m_\alpha O_{LL} + m_\beta O_{RR})\overline{g}(t) + m_E O_{LR} \overline{h}(t)\right) \\
O_{LL} &= L^*_{\beta i} L_{\alpha i} \quad O_{RR} = R^*_{\beta i} R_{\alpha i} \nonumber \\ 
O_{LR} &= L^*_{\beta i} R_{\alpha i} \quad O_{RL} = R^*_{\beta i} L_{\alpha i} \nonumber \\
g(t) &= \frac{i}{192\pi^2 m^2_{H_i} } \left[\frac{(t-1)(t^2 - 5t + 20) + 6t\ln t}{(t-1)^4}\right] \nonumber \\
h(t) &= \frac{i}{32\pi^2 m^2_{H_i}} \left[\frac{t^2 - 4t + 3 + 2\ln t}{(t-1)^3}\right] \nonumber \\
\overline{g}(t) &= \frac{i}{192\pi^2 m^2_{H_i} } \left[\frac{(t-1)(2t^2 + 5t - 1) + 6t^2(2t-1)\ln t}{(t-1)^4}\right] \nonumber \\
\overline{h}(t) &= \frac{i}{32\pi^2 m^2_{H_i}} \left[\frac{t^2 - 1 - 2t\ln t}{(t-1)^3}\right] \nonumber 
\end{align}
where $Q_{B,F}$ are the electromagnetic charges of the internal boson and fermion fields respectively and $t= (M_E/m_{\phi})^2$ with $M_E, m_{\phi}$ being the masses of internal fermion and scalar respectively.
\begin{figure}[!h]
\centering
\epsfig{file=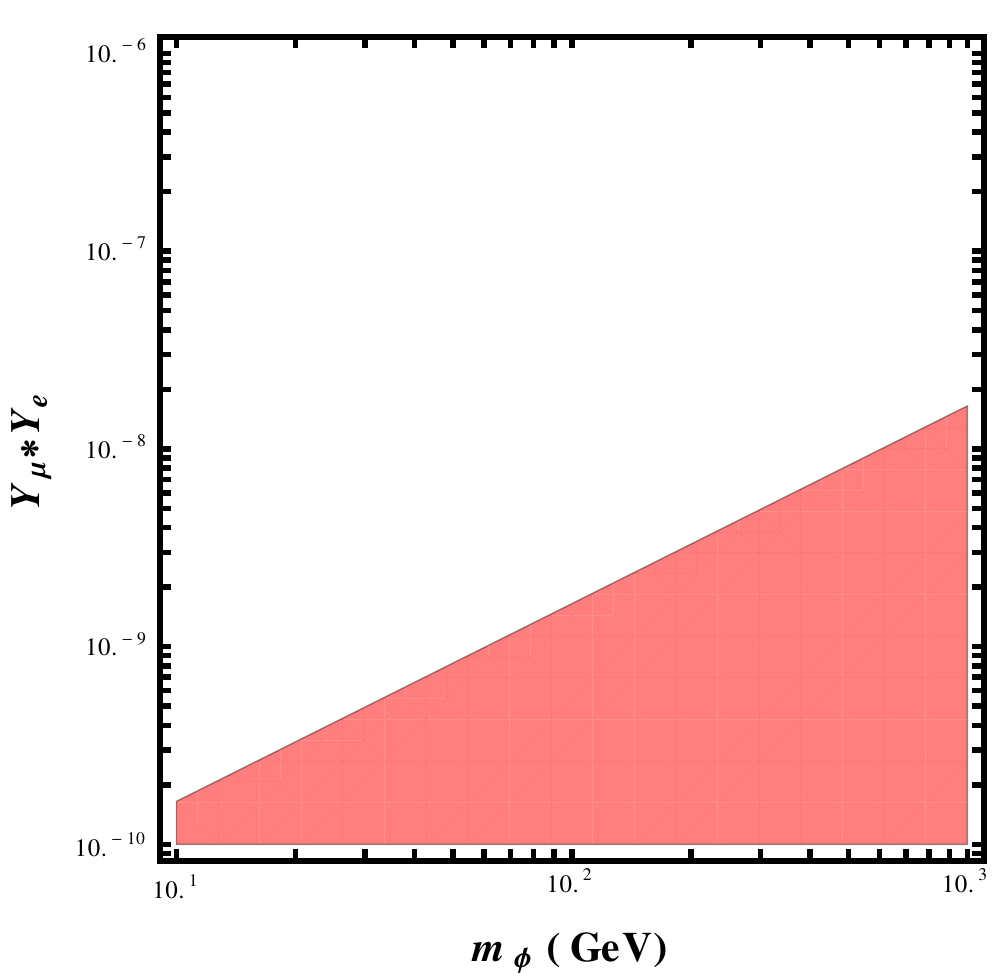,width=0.75\textwidth,clip=}
\caption{Parameter space in terms of Yukawa couplings and intermediate scalar mass $m_{\phi} = M_E/10$ such that $\text{BR}(\mu \rightarrow e \gamma) < 4.2 \times 10^{-13}$}
\label{fig6}
\end{figure}
\begin{figure}[!h]
\centering
\epsfig{file=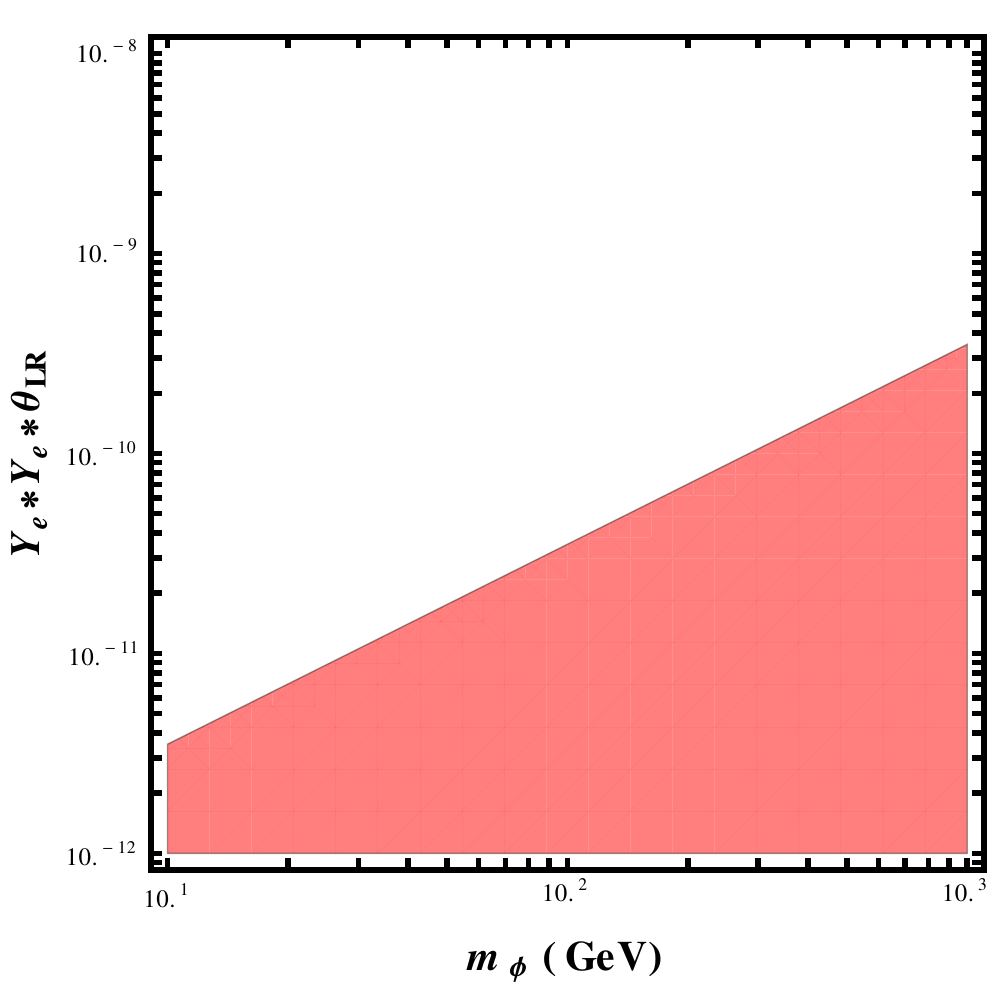,width=0.75\textwidth,clip=}
\caption{Parameter space in terms of Yukawa couplings and intermediate scalar mass $m_{\phi} = M_E/10$ such that $\lvert d_e \rvert/e < 8.7 \times 10^{-29} \; \text{cm}$. In the y axis, the $\theta_{LR}$ is the phase that arise from the mixing in the scalar sector.}
\label{fig7}
\end{figure}
\begin{figure}[!h]
\centering
\epsfig{file=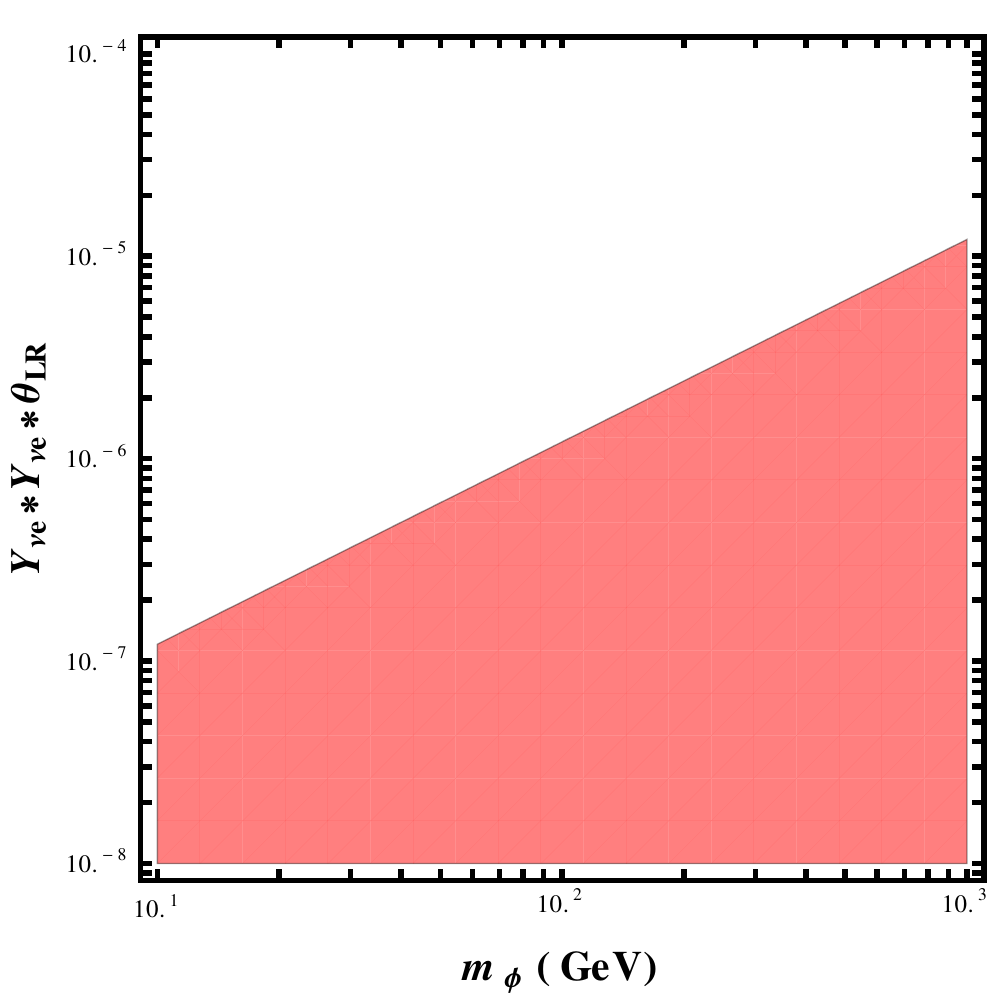,width=0.75\textwidth,clip=}
\caption{Parameter space in terms of Yukawa couplings and intermediate scalar mass $m_{\phi} = M_E/10$ such that $\lvert d_{\nu} \rvert/e < 3 \times 10^{-24} \; \text{cm}$. In the y axis, the $\theta_{LR}$ is the phase that arise from the mixing in the scalar sector.}
\label{fig8}
\end{figure}

Similarly one can have new contributions to the electric dipole moment of both charged leptons and neutrinos as seen from figure \ref{fig4} and \ref{fig5}, \ref{fig5a} respectively. Unlike $\mu \rightarrow e \gamma$ discussed above, the EDM of leptons is a flavour conserving observable which is a measure of the coupling of the particle's spin to an external electric field. Similar to charged lepton flavour violating decays, lepton EDM's are also vanishingly small in the SM, far beyond experimental reach. For example, the electron EDM is generated only at four loop in the SM giving rise to a tiny value $\lvert d_e \rvert/e \sim 3 \times 10^{-38}$ cm \cite{EDM1} which is way below the current experimental limits on the EDM of charged leptons:
\begin{equation}
\lvert d_e \rvert/e < 8.7 \times 10^{-29} \; \text{cm} \;\;\;\; (\text{ACME})
\end{equation}
\begin{equation}
\lvert d_{\mu} \rvert/e < 1.9 \times 10^{-19} \; \text{cm} \;\;\;\; (\text{Muon} \;g-2)
\end{equation}
\begin{equation}
\lvert \text{Re}(d_{\tau}) \rvert/e < 4.5 \times 10^{-17} \; \text{cm} \;\;\;\; (\text{Belle})
\end{equation}
\begin{equation}
\lvert \text{Re}(d_{\tau}) \rvert/e < 2.5 \times 10^{-17} \; \text{cm} \;\;\;\; (\text{Belle})
\end{equation}
which have been measured by the ACME collaboration \cite{EDM2}, the Muon $(g-2)$ collaboration \cite{EDM3} and the Belle collaboration \cite{EDM4} respectively. In the present model, the lepton EDM can occur at one loop instead of four loop in the SM, and hence there is a chance of generating a value close to these experimental bounds. Similar to the calculation of LFV decay above, the EDM of leptons can be calculated as
\begin{align}
d_l &= 2 \Im \left(\sigma_R - \sigma_L\right)
\end{align}
where $\sigma_{L,R}$ are given by the same expressions as above.

Similar to the charged leptons, the neutrinos can also have EDM due to similar one loop diagrams shown in figure \ref{fig5}, \ref{fig5a}. The bound on tau neutrino EDM has been derived from the LEP data and found to be \cite{EDMnu1}
\begin{equation}
\lvert d_{\nu_{\tau}} \rvert/e <  2 \times 10^{-17} \; \text{cm} 
\end{equation}
On the other hand, from naturalness arguments a much more stronger bound on neutrino EDM was found to be \cite{EDMnu2}
\begin{equation}
\lvert d_{\nu_e}, d_{\nu_{\mu}}, d_{\nu_{\tau}} \rvert/e <  3 \times 10^{-24} \; \text{cm} 
\end{equation}
which is also close to the cosmological bound $ \lvert d_{\nu} \rvert/e < 2.5 \times 10^{-22} \; \text{cm}$ \cite{EDMnu3}.

Using the analytical expressions derived above and incorporating the experimental bounds, we then show the allowed parameter space in terms of the Yukawa couplings and the masses of intermediate scalar field $m_{\phi}$. For simplicity, we assume the vector like lepton masses to be $M_E = 10 m_{\phi}$. Larger value of $M_E$ is in fact necessary in the scotogenic version of the model, in order to ensure that the lightest $Z_2$-odd particle is the neutral component of $\eta_{L}$ or $\eta_R$. The allowed parameter space from satisfying the experimental bound on $\text{BR}(\mu \rightarrow e \gamma) $ is shown in figure \ref{fig6}. Similarly, the corresponding parameter space from the requirement of satisfying experimental bounds on electron and neutrino EDM's are shown in figure \ref{fig7} and \ref{fig8} respectively. It can be seen from these three figures that the electron EDM constraints put the strongest bounds on the parameters.

\section{Conclusion}
\label{sec4}
We have studied two minimal versions of left-right symmetric model that can generate tiny Dirac neutrino masses at one loop level: one with and one without a stable dark matter candidate. The interesting feature of the model is that, we do not need additional discrete symmetries to generate tiny Dirac neutrino masses in the model without dark matter. The charged fermion masses are generated from a common universal seesaw mechanism due to the presence of additional heavy vector like charged fermions, singlet under the $SU(2)_{L,R}$ gauge symmetry. The same vector like heavy charged leptons play the role of generating tiny Dirac neutrino masses at one loop level without requiring additional heavy neutral fermions. The model can be suitably extended by an additional discrete $Z_2$ symmetry to accommodate a stable dark matter candidate: the lightest neutral component of the $Z_2$ odd scalar doublets $\eta_{L,R}$. After discussing the details of both the models and the particle spectra, including neutrino masses, we briefly discuss the possible implications for cosmology as the model contributes additional relativistic degrees of freedom $N_{\text{eff}}$ in terms of right handed part of the light Dirac neutrino and also predicts a stable scalar dark matter candidate. We find that the Planck upper bound on the effective number of relativistic degrees of freedom keeps the gauge interactions of the right handed neutrinos very suppressed resulting in a lower bound on the respective gauge boson mass $M_{W_R} > 3.548$ TeV. This bound on the other hand, can dictate the mass of right scalar dark matter if it has to give rise to the correct dark matter relic abundance. Therefore, the scotogenic version of the model can indirectly relate the additional contribution to $N_{\text{eff}}$ from the right handed neutrinos with the dark matter relic abundance.

We then discuss the possible implications of the model in flavour physics, particularly focussing on charged lepton flavour violating decay $\mu \rightarrow e \gamma$ and electric dipole moments of both charged leptons and Dirac neutrinos. We constrain the parameter space from the requirement of satisfying the current experimental bounds. 
Since the model can give sizeable contributions to LFV and EDM, near future experiments should be able to probe certain region of parameter space in the model. Also, since the model predicts vanishing amplitude for neutrinoless double beta decay, future experimental results in these experiments will also play a very crucial and decisive role as far as this model is concerned. The collider implications of such a model can also be very different from the conventional LRSM due to the absence of the usual bidoublet and triplet scalars. We intend to perform a detailed collider study of these scenarios in a subsequent work.

\bibliographystyle{JHEP}

\end{document}